\documentstyle[preprint,prc,eqsecnum,epsfig,aps,epsf]{revtex}

\begin{document}
\bibliographystyle{/usr/share/texmf/tex/latex/revtex/prsty}
\draft
%\wideabs{
\title{Lorentz Boosted NN Potential for Few-Body Systems: \\
       Application to the three-nucleon bound state}

\author{H. Kamada$^1$
\footnote{email:kamada@mns.kyutech.ac.jp},
W. Gl\"ockle$^2$\footnote{
email:walter.gloeckle@tp2.ruhr-uni-bochum.de},
 J. Golak$^{2,3}$\footnote{email:ufgolak@cyf-kr.edu.pl}, and
  Ch. Elster$^4$\footnote{email:elster@ohiou.edu}
}
\address{$^1$
Department  of Physics, Faculty of Engineering, Kyushu Institute of Technology,
 Kitakyushu 804-8550, Japan}

\address{$^2$
Institut f\"ur  Theoretische Physik II,
         Ruhr-Universit\"at Bochum, 44780 Bochum, Germany}

\address{$^3$ M. Smoluchowski Institute of Physics, Jagiellonian University, PL-30059
         Krak\'ow, Poland}

\address{$^4$ Institute of Nuclear and Particle Physics, and
Department of Physics, Ohio University, Athens, OH 45701, USA}

\date{\today}

\maketitle

\begin{abstract}
A Lorentz boosted two-nucleon potential is introduced in the context
of equal time relativistic quantum mechanics.
The dynamical input for the boosted nucleon-nucleon (NN) potential 
is based on realistic NN potentials, which by a suitable scaling of the momenta are 
transformed into NN potentials belonging to a relativistic
two-nucleon Schr\"odinger equation in the c.m. system.
This resulting Lorentz boosted potential is consistent with a previously 
introduced boosted two-body $t$-matrix.
It is applied in relativistic Faddeev equations for the three-nucleon bound state to
calculate the $^3$H binding energy.
Like in previous calculations 
the boost effects for the two-body subsystems are repulsive 
and lower the binding energy.
\end{abstract}

\narrowtext

\section{Introduction}
 Modern realistic nucleon-nucleon (NN) 
potentials using  a sufficiently large number of parameters
describe current NN phase shifts very well. The most prominent ones 
are CD-Bonn\cite{CDBONN}, Nijmegen 93, I,II \cite{nijm93},
and Argonne AV18~\cite{av18}. They predict NN observables up to about 350 MeV
nucleon laboratory energy perfectly well with a $\chi ^2 / N_{data} \sim 1$. 
This potential description is linked to a nonrelativistic Schr\"odinger equation.
Converged, nonrelativistic three-nucleon bound state calculations based upon these
potentials give values for the triton binding energy between 8.0 and 7.6~MeV
\cite{Friar,anbench,NoggaPhd}, whereas the
experimental result is 8.48~MeV. There is some sensitivity to nonlocalities in the NN
interactions, which influences the value of the triton binding energy. 
While the Nijmegen 93 and I and the AV18 potential are local, Nijmegen II and CD-Bonn
incorporate nonlocalities either via a $p^2$ dependence or the nonlocality structure given
by Dirac spinors and so-called `minimal relativity' factors. 
It is well established by now that the purely local interactions in the
above list lead to a triton binding energy of about 7.7~MeV, leaving about 0.8~MeV
unexplained. 

In general, the missing binding energy is attributed to some combination of nonlocality in
the NN interactions, three-nucleon force effects and relativistic effects. Of course,
these effects are often intermingled, i.e. relativity can motivate specific nonlocalities,
and negative energy components of a nucleon wave function could be viewed as a specific
subset of three-nucleon forces \cite{Coester1}.

The estimation of relativistic effects on the binding of three nucleons has been the
focus of a lot of work. However, up to now there has not been reached closure even on the sign of

a relativistic contribution to the three-nucleon binding energy. 
There are essentially two different approaches to a relativistic three-nucleon bound state
calculation, one is a manifestly covariant scheme linked to a field theoretical approach,
the other one is based on relativistic quantum mechanics on space-like hypersurfaces
(including the light front) in Minkowski space.
Within the first scheme Rupp and Tjon \cite{Rupp} find attractive corrections to the
triton binding energy, using separable interactions to facilitate the solution of the
Bethe-Salpeter-Faddeev equations. A three-dimensional reduction of this equation
also finds attractive contributions \cite{Sammarruca}.  The calculations of Stadler 
{ \it et al.} \cite{Stadler}
are based on a relativistic three-nucleon equation, incorporate the effects of Dirac
spinors and include negative energy components. They also incorporate off-shell effects of
the negative energy components of the Dirac spinors, and by varying this part, which is
not constraint by on-shell NN data,
they can achieve an attractive contribution to fit the experimental value. 
Within the second scheme  the relativistic
Hamiltonian consists of relativistic kinetic energies, two- and
many-body interactions including their boost corrections. The boost
corrections are dictated  by the Poincar\'e
algebra \cite{Bakamjian,Foldy,Keister}.  There exist already  applications
for the three-nucleon bound state \cite{Carlson,Gl86}, which suggest  a
repulsive contribution to the three-nucleon binding.
Thus,
the relativistic effects found in the two schemes appear 
controversial, in the approach based on field theory,  relativistic effects increase the
triton binding energy,  in the approach based on  
relativistic Hamiltonians, relativistic effects decrease the triton binding energy.

To the best of our knowledge we are not aware of any three-nucleon (3N) scattering
calculation including relativity in one or the other scheme due to the increased
difficulty of a scattering calculation.  In order to extend the
Hamiltonian scheme in equal time formulation to 3N scattering it would be a very
convenient starting point to have the Lorentz boosted NN potential which generates
the NN $t$-matrix in a moving frame via a standard Lippmann-Schwinger equation. 
In this paper we work out the NN potential in an arbitrary frame, and thus place our
work in the scheme based on relativistic quantum mechanics.
As application of our Lorentz-boosted potential we restrict ourselves to the
calculation of the triton binding energy.

The starting point for an NN potential in an arbitrary moving frame
is the interaction in the
two-nucleon c.m. system, which enters  a relativistic NN Schr\"odinger or
Lippmann-Schwinger 
equation. While realistic NN potentials are defined and fitted in the context 
of the nonrelativistic Schr\"odinger equation, NN potentials
refitted with the same accuracy in the framework of the relativistic NN Schr\"odinger 
equation do not yet exist. A first step in that direction has been done 
in Ref.~\cite{Carlson}, where  the AV18 potential has been
refitted to describe NN phase shifts in the relativistic context.
Here we prefer to use a different route and
employ an analytical scale transformation 
of momenta which relates NN potentials in the nonrelativistic and relativistic 
Schr\"odinger equations such that
exactly the same NN phase shifts are obtained by both equations \cite{Kamada}. 
Though this transformation is not a substitute for  a NN potential 
with proper relativistic features and though it suffers from
some conceptual defects~\cite{Polyzou}, it can serve  the purpose of this
work, namely to 
 illustrate the effects of a  Lorentz boosts on NN potentials.

This paper is organized as follows. 
In Section II we derive and explicitly formulate 
the Lorentz-boosted potential related to a given nonrelativistic potential.
In Section III we apply our formulation to the Reid Soft Core potential and discuss
features of the Lorentz boost. Then we solve
the relativistic 
3N Faddeev equation based on that  Lorentz-boosted potential.
We end with a brief summary and outlook in section IV.

\section{Derivation of the Boosted Potential}

A formalism for treating the relativistic three-body Faddeev equations has 
been introduced in \cite{Gl86}. 
There a  Lorentz boosted t-matrix  is constructed from  the 
relativistic two-nucleon 
$t$-matrix given in  the nucleon-nucleon (NN) center-of-mass (c.m.) system. 
The  NN t-matrix obeys the relativistic 
Lippmann-Schwinger equation:
\begin{eqnarray}
t(\vec k , \vec k_0) = v (\vec k, \vec k_0)+ \int d^3 k' {{v(\vec k,\vec k') ~t(\vec k',\vec k_0) }
\over { \omega (\vec k_0)-\omega  ( \vec k') + i\epsilon }},
\label{R-LS}
\end{eqnarray}
where $v(\vec k,\vec k')$ is the relativistic potential given in the c.m. system 
with $\vec k$ and $-\vec k$ the individual
momenta in that system 
and $\omega(\vec k) =2  \sqrt{ k^2 + m^2}$.
In Ref.~\cite{Gl86} 
a boosted potential two-nucleon potential $V$ (Eq. (3.4) of \cite{Gl86}) is naturally 
introduced as
\begin{eqnarray}
\label{Vrel}
V \equiv \sqrt{ [\omega (\vec k) + v]^2 +{ \vec p~}^2 }
-\sqrt{ \omega (\vec k)^2 +  {\vec p~}^2   } ,  
\end{eqnarray}
where $\vec p$ is the total momentum of the two-nucleon system.
Obviously, for $\vec p = 0$ one obtains $V=v$.
For any  application to the three-nucleon system, one needs to be able to calculate the
matrix elements of $V$ explicitly. Thus  
we need an explicit representation and 
use a momentum space form based on eigenstates of the c.m. momentum operator 
$\vec k$ to obtain the matrix elements
\begin{equation}
\langle \vec k \vert V (\vec p) \vert \vec k' \rangle     = \langle \vec k \vert  
\sqrt{ [\omega ( k) + v]^2 +{  p}^2} 
 \vert \vec k' \rangle
-\langle \vec k \vert 
\sqrt{ \omega (k)^2 +  { p}^2  } \vert \vec k' \rangle .
\label{eq:3}
\end{equation}
With the help of the eigenstates $\psi _b$ and 
$ \vert \vec k \rangle ^{(+)} $ (bound and scattering eigenstates) 
of the c.m. Hamiltonian ($\omega (k) + v$) the completeness relation is given as
\begin{eqnarray}
{\bf 1} = 
\vert \psi_b  \rangle    \langle \psi_b \vert +  \int  \vert \vec k \rangle ^{(+)}
~d^3 k ~ ^{(+)} \langle \vec k  \vert.
\label{One} 
\end{eqnarray}
Inserting the completeness relation into Eq.~(\ref{eq:3}) leads to
\begin{eqnarray}
\label{eq:5}
 \langle \vec k \vert V (\vec p) \vert \vec k' \rangle & =&   \\
& &\langle \vec k \vert \psi _b \rangle \langle \psi _b \vert \vec k'\rangle \sqrt{M_b^2 +p^2}
+ \int d^3 k'' \langle \vec k \vert \vec k'' \rangle^{(+)}\sqrt{\omega (k'')^2+ \vec p^2}
 \; ^{(+)} \langle \vec k '' \vert \vec k' \rangle \cr
 & & -   \delta ( \vec k - \vec k') \sqrt{\omega (k)^2+\vec p ^2}, \nonumber
\end{eqnarray}
where $M_b$ is the bound state mass. 
Using the  standard relation between scattering states and plane wave states
\begin{eqnarray}
 \vert \vec k \rangle ^{(+)} = \vert \vec k  \rangle  + G_0 ^{(+)} t \vert
 \vec k  \rangle , 
\end{eqnarray}
the potential matrix element from Eq.~(\ref{eq:5}) can be rewritten as
\begin{eqnarray}
\label{eq:7}
\lefteqn{\langle \vec k \vert V (\vec p) \vert \vec k' \rangle } \cr
&=& \psi_b (\vec k)\; \sqrt{M_b^2+p^2} \; \psi_b (\vec k') \cr
& & + {{ t ^* (\vec k', \vec k; \omega ) }\over {\omega -\omega' -i \epsilon} }
\sqrt{ \omega ^2 + p^2}  + 
{{ t(\vec k ,\vec k';\omega ') } \over {\omega' -\omega +i\epsilon}} \sqrt{{\omega'}^2+p^2} 
\cr 
& & + 
\int d ^3 k'' {{ t(\vec k ,\vec k'';\omega '') } \over {\omega'' -\omega +i\epsilon}} 
\; \sqrt{{\omega''} ^2+p^2} \; {{ t ^* (\vec k', \vec k''; \omega'' ) }
\over {\omega'' -\omega' -i \epsilon} }
\cr & & \cr
 &=& v(\vec k, \vec k') + \psi_b(\vec k) ( \sqrt{M_b ^2 + p^2} - M_b ) \psi_b (\vec k')
\cr 
& & + {1 \over {\omega-\omega'}} \biggm[ (\sqrt{\omega^2 +p^2} -\omega)\;
 \Re[ t(\vec k',\vec k;\omega)]
 - (\sqrt{\omega'^2 +p^2} -\omega') \; \Re[t(\vec k,\vec k';\omega ')] \biggm]
\cr & & +  {1 \over { \omega - \omega '} }  
 \Biggm[ 
  {\cal P }  \int d^3 k'' { (\sqrt{{\omega''} ^2 +p^2} -\omega'') \over {{\omega'' -\omega}} }
t (\vec k , \vec k'' ; \omega'') t^{*} (\vec k' , \vec k'' ; \omega'') 
\cr & & - {\cal P } \int d^3 k''{ (\sqrt{{\omega''} ^2 +p^2} -\omega'' ) \over {\omega'' -\omega'}} 
t (\vec k , \vec k'' ; \omega'') t^{*} (\vec k' , \vec k'' ; \omega'') \Biggm] .
\end{eqnarray}
Here ${\cal P}$ denotes the principal value prescription,
and  $\omega ' = 2\sqrt{{k'}^2+m^2} $, and $\omega '' = 2\sqrt{{k''}^2+m^2}$.
Note that the matrix elements is well defined for $ \omega = \omega '$, since both 
brackets vanish in this case.

Thus, the boosted potential, which  depends on the total two-nucleon momentum $\vec
p$, requires the knowledge of the NN bound state wave function
 and the half-shell NN $t$-matrices given 
in the 2N c.m. system. For any given potential  $v$ those quantities can be
calculated by standard methods. 
Once the matrix elements $\langle \vec k \vert V (\vec p) \vert \vec k' \rangle \equiv
V(\vec k , \vec k' ; p)$ are known, 
the boosted $t$-matrix elements $T \equiv T (\vec k , \vec k' ; p)$ 
can be calculated from the relativistic 
Lippmann-Schwinger equation,
\begin{equation}
T(\vec k , \vec {k'}; \vec p)= V(\vec k , \vec {k'}; \vec p)
+
 \int d^3 k'' { V(\vec k , \vec {k''}; \vec p) T(\vec k'' , \vec {k'}; \vec p) \over
{\sqrt{ { \omega' } ^2+p^2} - \sqrt{{\omega''}^2+p^2} +i \epsilon }}.
\label{eq:8}
\end{equation}
For any given  momentum $\vec p$ the bound state wave function $ \psi_b $ 
defined in the c.m. system obeys 
\begin{eqnarray}
\psi_b (\vec k )= { 1\over {\sqrt{M_b^2+p^2} - \sqrt{ { \omega }
^2+p^2} }  }   \int d^3 k' V (\vec k, \vec { k'}; \vec p ) 
\psi_b (\vec  { k'} ).
\end{eqnarray}
This eigenvalue equation is an excellent numerical test for the numerical calculation
of the boosted potential, since it
has to reproduce exactly the boosted energy  $ \sqrt{ M_b^2 + p^2} $ 
of a deuteron in motion.

In Ref.~\cite{Gl86} the Lorentz boosted $T$-matrix was already introduced but calculated in
a different fashion (see Eq.(3.27) therein). In the procedure we suggest here, we want
to focus on the calculation of the  boosted potential. We want to remark that
despite the occurrence of complex valued half-shell t-matrices in Eq.~(\ref{eq:7}) 
the potential matrix element $V(\vec k , \vec {k''}; \vec p)$ is real. The proof is
given in Appendix A. As a consequence of that and the manifest symmetry of 
$\langle \vec k \vert V (\vec p) \vert \vec k' \rangle $ 
it folows from Eq.~(\ref{eq:8}) that one can define an unitary S-matrix.
We define the S-matrix as              
\begin{eqnarray}
S(\vec k ,\vec { k'}; \vec p ) \equiv { 4 \over { k \sqrt{ \omega ^2 + p^2 }}}
\; \delta ( \sqrt{ \omega ^2 + p^2 } - \sqrt{ {\omega ' }^2 + p^2 } ) 
\hat S (\vec k ,\vec { k'}; \vec p),
\end{eqnarray}
with
 \begin{eqnarray}
 \hat S (\vec k ,\vec { k'}; \vec p ) \equiv \delta (\hat k - \hat {k'}) 
- 2i \pi { k \sqrt{ \omega ^2 + p^2 } \over 4 } \; T ( \vec k, \vec {k'} ; \vec p ).
\end{eqnarray}
Then one can show in the standard way that
\begin{eqnarray}
\int d \hat {k''} \hat S (\vec k ,\vec { k''}; \vec p )
 \hat S^* (\vec k' ,\vec { k''}; \vec p ) = \delta ( \hat  k - \hat {k'}). 
\end{eqnarray}
Note that $\vec k = k \hat k $, $\vec {k''} = k \hat {k''}$, and $\vec {k'} = k \hat {k'} $, 
where the quantities $\hat k$, $\hat {k'} $, and $\hat {k''}$ are unit vectors.

\section{Applications}

In this section we would like to consider the  Lorentz boosted
potential and use it to calculate the binding energy of the triton.

\subsection{Calculation of the boosted potential}

First we need  to construct a suitable  
potential which enters  the relativistic Lippmann
Schwinger equation,  Eq.~(\ref{R-LS}).
A standard  nonrelativistic potential fulfills the
nonrelativistic Lippmann Schwinger equation for the nonrelativistic 
t-matrix $t^{(nr)}$,
\begin{eqnarray}
\label{eq:3.1}
t^{(nr)}(\vec q , \vec { q'}) = v^{(nr)} (\vec q, \vec { q'} )+ \int d^3
q '' { {  v^{(nr)} (\vec q, \vec {q''}) \; t^{(nr)}(\vec {q ''},\vec {q '} ) }
\over { { \vec {  q ' } ^2 \over m}  - {{  \vec {q ''} }^2 \over m   } +
 i\epsilon }}.
\end{eqnarray}
Using a relativistic propagator in Eq.~(\ref{eq:3.1})  will not result in the same phase
shifts or observables. However, there is a scale transformation, which generates a phase
equivalent relativistic potential $v$ from a nonrelativistic
potential $v^{(nr)}$~\cite{Kamada}. This scale transformation is derived from requiring
that
the  relativistic and nonrelativistic form of the kinetic energy give the same result.
This requirement leads to analytic relations connecting the nonrelativistic momentum
$q$ with the corresponding relativistic momentum $k$, 
\begin{equation}
q \buildrel ! \over \equiv \sqrt{m} \sqrt{\omega ({ \vec k}) - 2m
 }.\label{scale}
\end{equation}
The details of the derivation are
given in Ref.~\cite{Kamada}. Here we only list the results necessary for the understanding
of our present
considerations. For a given nonrelativistic potential $v^{(nr)}(\vec q, \vec { q' })$ the
corresponding phase equivalent relativistic potential is obtained as
\begin{equation}
v(\vec k,\vec{ k' })={1 \over h(q)  } v^{(nr)}(\vec q, \vec { q' }){ 1 \over h(q')  }.
\label{vrel}
\end{equation}
The corresponding relativistic NN t-matrix can be obtained from the nonrelativistic one in
a similar fashion, 
\begin{equation}
t(\vec k, \vec{ k' };\omega ')= {1 \over h(q)  }~t^{(nr)} (\vec q, \vec { q' })
 { 1 \over h(q')  }.
\label{trel}
\end{equation}
The Jacobian function $h(q)$ is defined as   
\begin{eqnarray}
h(q) \equiv   \sqrt{ ( 1 + { q^2 \over 2 m ^2 } ) \sqrt{ 1 + { q^2 \over   4  m ^2 } } }.
\end{eqnarray}
The results of Eqs.(\ref{vrel}) and (\ref{trel}) can be derived from  the scale transformation 
of Eq.~(\ref{scale}). Of course, they are not 
equivalent to the introduction of a relativistic
potential $v$ and a corresponding NN t-matrix based on a field theory
theory. However,
for our purposes, the scale transformation is a very useful and simple parameterization
of a relativistic NN potential,  which conserve the NN phase shifts exactly, and which can
enter Eq.~(\ref{Vrel}) for the boosted potential. We want to remark here, that
Eq.~(\ref{Vrel}) is general and  independent of the way, the relativistic potential
was obtained. 

In order to study the effect of the boost on the potential in more detail, we choose
as the nonrelativistic potential the Reid soft core potential
 (RSC) \cite{RSC}. The RSC potential in the $^1S_0$ state  is given by
\begin{eqnarray}
v^{(nr)}(r)= (  -10.463 { e^{ - \mu r } \over \mu r } 
            -1650.6  { e^{ - 4\mu r } \over \mu r } 
             + 6484.2 { e^{ - 7\mu r } \over \mu r } )~~{ \rm MeV }
\label{RSCpot}
\end{eqnarray}
where $  \mu  $ is 0.7 fm$^{-1 }$. 
 
\begin{figure}[hbt]
\centerline{\psfig{file=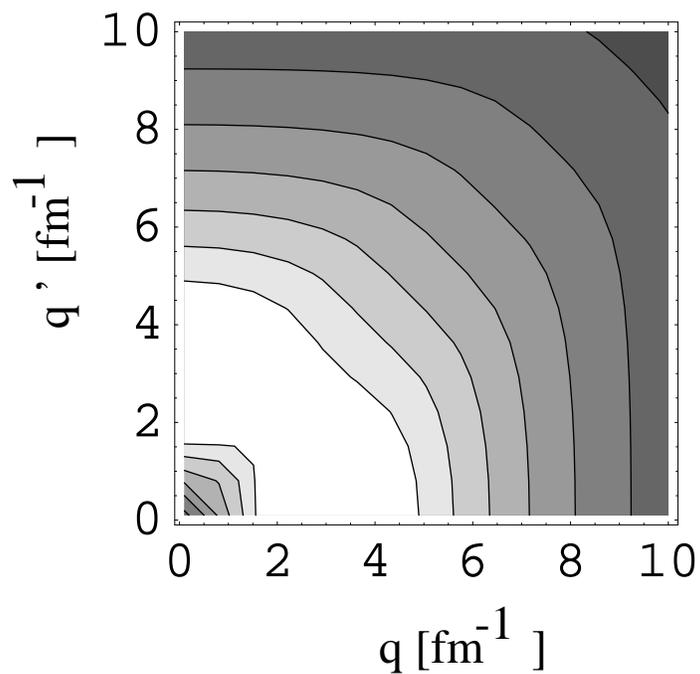,angle=0}}%,scale=1.}
\caption{Contour plots for the 
Reid soft core potential $v^{(nr)} (q,q')$ in  the  state
$^1S_0$  in momentum space. All values are
positive, decreasing from light to darker shades.}
\label{fig1}
\end{figure}

\begin{figure}[hbt]
\centerline{\psfig{file=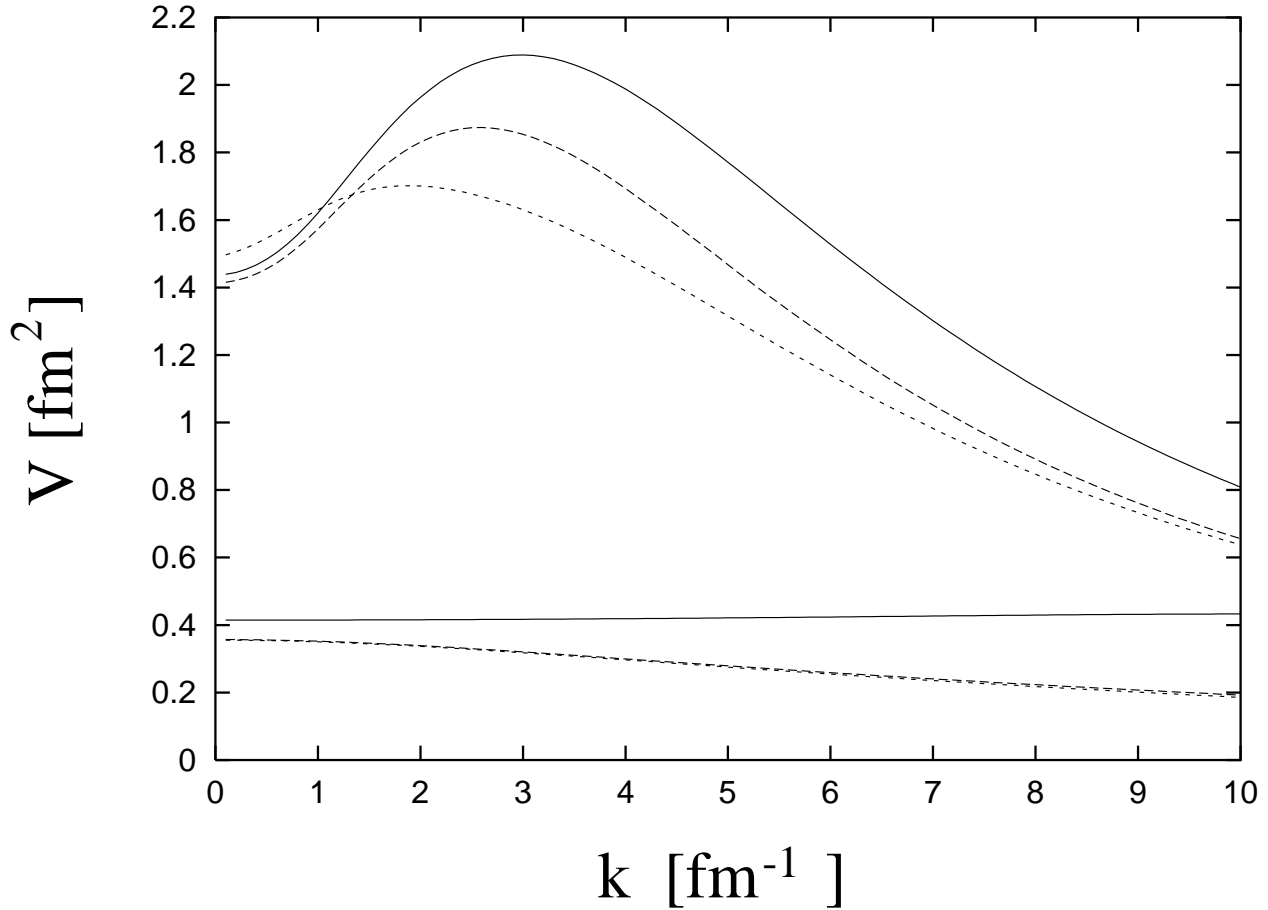,width=140mm,angle=-90}}
\caption{The nonrelativistic potential $v^{(nr)}(q,q')$ (solid line), 
the scale transformed potential
$v(k,k')$ (long dashed line) and the boosted potential $V(k,k';p)$ (short dashed line) 
as function of the momentum $k$ and two fixed momenta $k'$.  All potentials  are projected
on the $^1$S$_0$ partial wave state.
The figure shows  two groups of 
lines: the upper group ( V $>$ 0.6fm$^{-1}$ ) is calculated for a fixed 
$k'$=1~fm$^{-1}$ and the lower one for a fixed $k'$=15~fm$^{-1}$. 
The boosted potential $V(k,k';p)$  is evaluated at $p$=20fm$^{-1}$. 
 }
\label{fig2}
\end{figure}

In  Fig.~\ref{fig1}  a contour plot of $v^{(nr)}(q,q')$ is given for the $^1S_0$ state.
It should be noted that this particular potential is positive for all values of $q$ and
$q'$. Next, we successively apply first the scale transformation and then the boost 
to $v^{(nr)}(q,q')$. 
In Fig. \ref{fig2} we compare the projections on the $^1S_0$ state  of the 
three potential functions, namely $v^{(nr)} (\vec q, \vec q')$,
 $v(\vec  k, \vec k')$, and
$V( \vec k, \vec {k'}; \vec p)$ as a function of $ k$. 
Since the scale transformation of Eq.~(\ref{scale}) changes the momentum scale, we
express  $v^{(nr)} (\vec q, \vec q')$ in terms of $k$ and $k'$, in order to compare it with
the other two potential functions.
We choose two fixed values for $k'$, namely $k'$=1~fm$^{-1}$ and
$k'$=15 fm$^{-1}$. 
The total two-nucleon momentum is chosen as $p$=20 fm$^{-1}$. 
First, we would like to make some more general remarks. 
For small momenta $q$, i.e. in a very nonrelativistic regime, on has  $ q \approx$ $k$.
Furthermore, since the function $h(q)$ is always larger than 1, $v$ will be always smaller
than  $v^{(nr)}$. At larger momenta $q$ differs from $k$ and the relation between
$v$ and $v^{(nr)}$ depends in general on the shape of $v$. 
In our case $v$ is always smaller than $v^{(nr)}$. 
The boost effect leads  to another overall decrease 
of the values of $V$ except for small momenta, where $V$ is larger than $v$  and $v^{(nr)}$.
We also want to point out that we chose quite a large two-nucleon momentum $p$ in order to
show the effects of the boost. For two-nucleon momenta in the order of about 5~fm$^{-1}$,
the boost effects are much smaller. In fact, almost all of the difference between $v$ and
$v^{(nr)}$ would be given by the scale transformation, i.e. by an underlying different
scattering equation.

\subsection{Calculation of the triton binding energy}

 Now let us move on to entering the boosted NN potential into the relativistic
three-body Faddeev equation to calculate the bound state of $^3$H.
The relativistic Faddeev equation as already introduced in Ref.~\cite{Gl86}
reads
\begin{eqnarray}
\label{Faddeevrel}
\phi(\vec k,\vec p)=  { 1 \over E_b - {\cal E } (\vec k , \vec p) }  
\int d ^3 p ' { T_a ( \vec k , \vec \kappa ( \vec {p'}, - \vec p- \vec {p'}); \vec p)
\over {{\cal N }  ( \vec {p'}, -\vec p - \vec {p'}){\cal N }
( -\vec p -\vec {p'}, \vec p) }} \;
\phi(\vec \kappa (-\vec p - \vec {p'}, \vec p), \vec {p'} ) 
\end{eqnarray}
where $ \phi$ is the Faddeev component and $E_b$ the three-body binding energy.
The index $a$ at the boosted T-matrix indicates a properly antisymmetrized two-body
T-matrix.
The vector $\vec k$ represents the
relative momentum in the two-body c.m. subsystem as in the nonrelativistic case, and 
$\vec p$ stands for the
momentum of the corresponding third particle. At the same time $\vec p$ is the
(negative) total momentum of the two-body subsystem and thus is responsible
for the boosts of that subsystem. Clearly, the three individual nucleon momenta
sum up to zero. If $\vec p$ and $\vec {p'}$ are the momenta  of two individual
nucleons, then their relative
momentum (half the momentum difference) in their c.m. system is obtained through a 
Lorentz transformation, which is explicitly given as
\begin{eqnarray}
\vec \kappa (\vec {p}, \vec {p'}) \equiv { 1 \over 2 } 
\left[  \vec {p}  - \vec {p '}
-( \vec {p}  +\vec {p'} )
 {  \Omega  - \Omega' \over ( \Omega +\Omega' ) +
\sqrt{ (\Omega+ \Omega')^2 - (\vec {p} + \vec {p'})^2  } } \right],
\end{eqnarray}
  where $\Omega (p)=\sqrt{m^2 + {p }^2} =\Omega  $ and  $ \Omega' =\sqrt{ m^2 +{p'} ^2  }$.
The last term in Eq. (3.8) reflects the relativistic effect in the definition of a relative momentum.
When going from individual momenta $\vec {p}$ and $ \vec {p'}$  of the subsystem 
to the relative 
momentum and the total two-body momentum $ \vec { p} + \vec {p'}$ one has to consider
the Jacobian of that transformation. The square root of the Jacobian is given by
\cite{Gl86}
\begin{eqnarray}
\label{Norm}
{\cal N }( \vec {p}, \vec {p'}) &=& \left[ \left\vert {\partial (\vec {p}
 , \vec{ p'} ) \over \partial ( \vec k, \vec {p} + \vec {p '} )}
 \right\vert \right]^{1/2}  \cr 
&=& 
 \left[ 4\Omega \Omega' \over {\sqrt{ (\Omega+\Omega')^2 - (\vec {p} + \vec {p'})^2  }
( \Omega+ \Omega')  }  \right]^{1/2}.
\end{eqnarray}
The kinetic energy ${\cal E }$ is given by
\begin{eqnarray}
 {\cal E }( \vec k , \vec p ) &=&  \sqrt{ \omega^2 (k) + m^2 }+ \Omega
 -3m
\cr 
&=& \Omega + \Omega'+\Omega'' -3m. 
\label{kinetic}
\end{eqnarray}
A detailed derivation of the above relations is given in  Ref.~\cite{Gl86}.

In our calculation of the triton binding energy 
the relativistic Faddeev equation, Eq.~(\ref{Faddeevrel})
is solved in a partial wave  basis. The explicit representation of Eq.~(\ref{Faddeevrel})
in a partial wave decomposition is given in Appendix B.
Since we are here only interested to test the feasibility
of our approach,  
we only perform a 5-channel calculation at present.
This means we allow the NN
forces to act only in the states $^1S_0$ and $^3S_1-^3D_1$ 
(see e.g.  table 3.4 in \cite{TEXT}).
We want to point out that in contrast to a nonrelativistic calculation
not only the Faddeev component and the T-matrix depend on the angle
between $\vec p$ and $\vec {p'}$ 
but also the Jacobian $\cal N$.
In this first approach we ignore the Lorentz transformation of the
spins states. As NN potentials we employ the high-precision potentials
 CD-Bonn \cite{CDBONN}, NijmI,II, 93 \cite{nijm93}, and  AV18\cite{av18}, as well as the
Reid Soft Core potential \cite{RSC} and two different Yamaguchi \cite{Gibson} potentials.
For all potentials (with the exception of RSC) we use np forces only.
With those potentials given, our calculation proceeds as follows. First, we perform the
scale transformation of Eq.~(\ref{scale}) to obtain a phase equivalent potential obeying
the relativistic two-body Lippmann-Schwinger equation. Then we boost this potential and
solve for the relativistic, boosted T-matrix, which enters the relativistic Faddeev
equation, Eq.~(\ref{Faddeevrel}). This is in contrast to the approach given in
Ref.~\cite{Gl86} where a relativistic NN t-matrix in the NN c.m. frame was calculated first, and
this t-matrix was boosted to the obtain the T-matrix entering Eq.~(\ref{Faddeevrel}).
We also want to point out that the relativistic potential used in Ref.~\cite{Gl86} is only
approximately phase shift equivalent to the nonrelativistic one.
%In Fig. \ref{flowchart} we show a simple flow chart, 
%comparing our steps with the ones in Ref. \cite{Gl86}.

Our results for the relativistic Faddeev
calculations based on five channels 
are displayed in Table~\ref{table1}.  For comparison, we also list the binding energies
$E_b^{(nr)}$ obtained from a nonrelativistic five channel calculation.
We want to emphasize that the underlying relativistic and nonrelativistic NN forces are 
strictly phase equivalent and give the same deuteron binding energy.
Only under these conditions it is reasonable to pin down relativistic effects in the
triton binding energy. From Table~\ref{table1} we see that the difference between the
relativistic and nonrelativistic binding energies span a range of about 0.29-0.43~MeV.
The Yamaguchi potentials do not fall into this range.
From Table~\ref{table1} we can conclude that the relativistically calculated triton
binding energy is reduced in magnitude compared to the one calculated nonrelativistically.
A related investigation was carried out in Ref.~\cite{Carlson} based on the AV18
potential. There the nonrelativistic NN potential was augmented by relativistic
corrections of low orders following the work of \cite{Foldy} and was refitted to the
NN phase shifts. The final relativistic correction  to the 
binding energy of $^3$H given in Ref.~\cite{Carlson} is 0.33MeV, which is comparable to
our present findings.  It is interesting to notice, that in the case of the Yamaguchi
potentials, which are purely attractive, the relativistic, repulsive effect is weaker,
namely only about  0.2~MeV. This is presumably connected to the absence of short range
repulsive force components, i.e. high momentum components, which are presumably  mostly
affected by the relativistic effects. However, it will be difficult to provide general
arguments on the relative size of the relativistic effects under consideration, since they
most likely depend on the specific functional form of the potential. 
We also want to mention, that for nonrelativistic calculations the contributions of the 
higher partial waves in the two-body subsystem are attractive and range from about
0.04 to 0.24~MeV. 

In order to shed some more light on the different contributions to our relativistic
calculation, we want to expose the effect of the normalization factor separately.
To do so, we solve Eq.~(\ref{Faddeevrel}) under the assumption that $ {\cal N} =1$, which
is the nonrelativistic  limit of that  quantity. 
The resulting binding energies are listed in the last column of Table~\ref{table1}.
replaced by 1, which is the nonrelativistic limit of that  quantity.
We see that $\cal N$ gives a repulsive contribution in all cases.

\begin{figure}[hbt]
\psfig{file=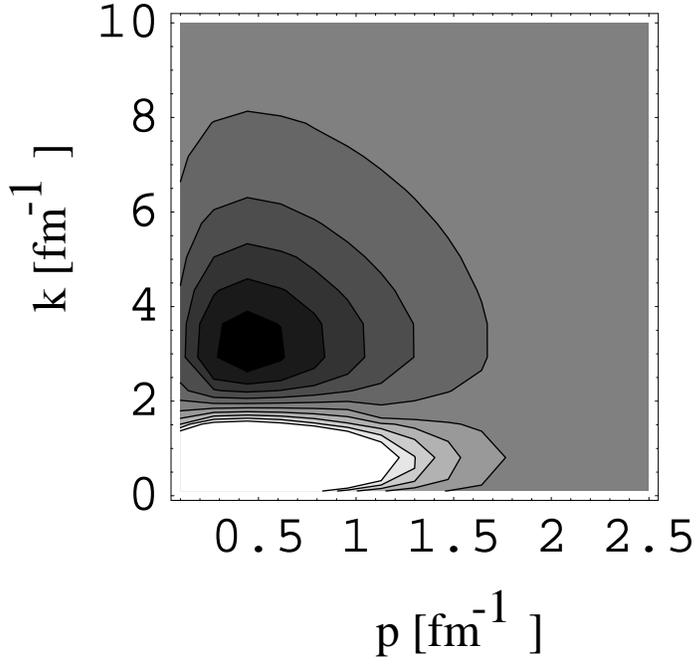,angle=0}%,scale=1.}
\caption{The relativistic  
Faddeev component $\phi (k,p)$ linked to the 
$^1$S$_0$ state in the 2-body subsystem. }
\label{fig4}
\end{figure}

\begin{figure}[hbt]
\psfig{file=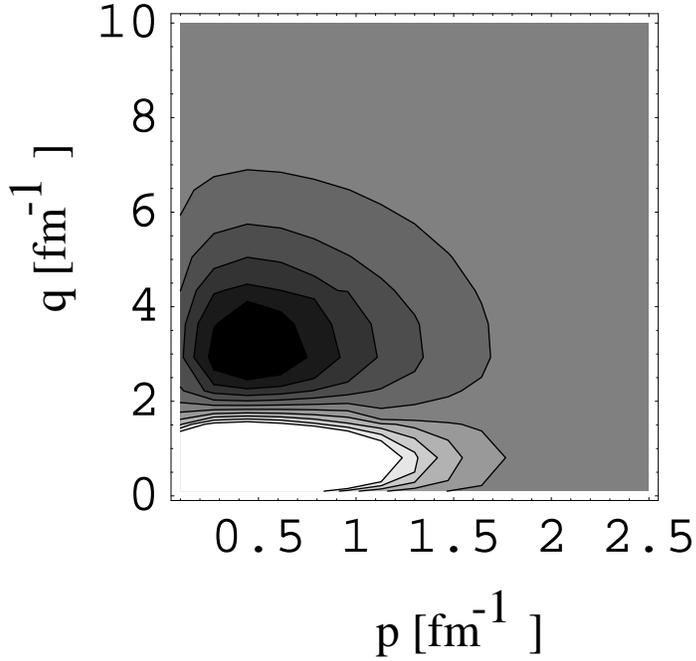,angle=0}
\caption{The nonrelativistic Faddeev component $\phi(q,p) $  corresponding to Fig. \ref{fig4}. 
The contour lines carry the same values as in Fig. \ref{fig4}. Note the difference in the 2-nucleon subsystem momentum  $q$ to $k$ in Fig. \ref{fig4}. }
\label{fig5}
\end{figure}

\begin{figure}[hbt]
\psfig{file=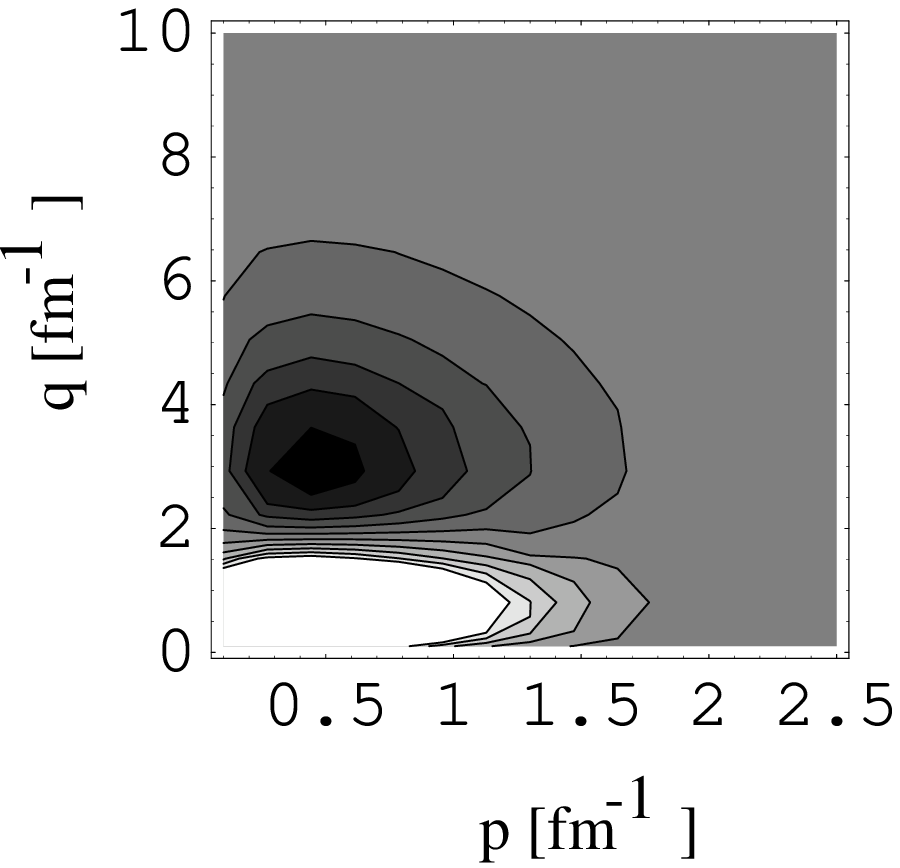,angle=0}
\caption{The relativistic  Faddeev component.  From Fig. \ref{fig4}, where $k$ is expressed in terms of $q$. 
Contour lines are as in Figs. \ref{fig4} and \ref{fig5}.  }
\label{fig6}
\end{figure}

Finally we display the relativistic and nonrelativistic Faddeev
components  in Figs.~\ref{fig4}  and \ref{fig5}. We choose the
channel related to the two-body state $^1S_0$, which  is one of the
five channels. The figures show that the relativistic Faddeev component 
is more extended into the
high $k$-region than the nonrelativistic one. As a reminder,  the corresponding
two-body relative momentum is denoted by $q$, see Eq.~(\ref{scale}).
However, when the momentum $k$ of 
the relativistic Faddeev component is expressed in terms
of $q$ according to Eq.~(\ref{scale})  and the component is replotted as a function of $q$,
then the shape of this Faddeev component is very close to the  nonrelativistic one, as
shown   
in Fig. \ref{fig6}.

\section{Summary and Outlook}

We derived and presented  an explicit expression for a Lorentz boosted NN potential,
which can be used to determine the two-body T- matrix in a frame, in which the total
momentum of the two particles is different from zero. 
The description of two-body systems with non-zero total momentum is relevant for
calculating properties in an interacting three-body system in a relativistic framework. 
The
general T-matrix was inserted into a relativistic three-body Faddeev equation for the
bound state, which was
proposed in \cite{Gl86}. The dynamical input consisted of NN potentials
used in a relativistic two-body Schr\"odinger equation which are exactly
phase equivalent to nonrelativistic NN potentials used in the nonrelativistic
Schr\"odinger equation. The phase equivalence of the two different NN potentials is
achieved  by a momentum scale transformation \cite{Kamada}. 
We applied this scheme to various modern high precision NN
potentials and compared resulting three-nucleon binding energies from the
nonrelativistic and relativistic 3N Faddeev equations. In all case the
relativistic effects turned out to be repulsive and of the order of 400 keV. 

The effect of the boost turns out to be relatively small for moderate total momenta of the
two nucleons, however at high momenta they are quite visible.
If one compares the relativistic and nonrelativistic Faddeev components one
notices some enhancement for high momentum components in the two-body
subsystem.

The access to  boosted NN potentials opens the door to considering the
relativistic Faddeev equations for three-nucleon  scattering. The need for a
relativistic description of three-nucleon scattering became already apparent when
measurements of the
total cross section for neutron-deuteron scattering \cite{Abfalterer}
 were analyzed  within
the framework of nonrelativistic Faddeev calculations \cite{Witala99}. 
Here, NN forces alone were not
sufficient to describe the data above about 100 MeV. The discrepancy is most likely due to
missing corrections from three-nucleon forces and relativistic effects. The relativistic
corrections considered in Ref.~\cite{Witala99} were only of kinematic nature, but they
lead to an increase of the total cross section by about 3\% at 100~MeV and about 7\% at
250~MeV. This estimate, though very crude, emphasizes the importance of a
consistent treatment of relativistic effects especially in scattering. The
availability of a boosted NN potential is one step in that direction.
 Additional technical
steps in relation to the relativistic free 3-body propagator and its
singularity structure have already been worked out \cite{Kamada2000}. We expect that
the Wigner rotations of the spin states can be performed along the line given
in \cite{Keister}.

\vfill

\acknowledgements

This work is partialy supported by the U. S. Department of Energy under 
contract
No. DE-FG02-93ER40756 with Ohio University.
The numerical calculation have been performed on the Cray T3E and T90 
of the Neumann Institute for Computing (NIC) at the Forschungszentrum J\"ulich, Germany.

\appendix

\section{Proof of the Reality of the boosted Potential} 

The boosted potential is given in Eq.~(\ref{eq:8}). Obviously, the first three terms are
real. Here we show that the remaining forth term,
\begin{eqnarray}
  {1 \over { \omega - \omega '} } 
& \{  & {\cal P}  \int d^3 k''  {  (\sqrt{{\omega''} ^2 +p^2} -\omega'' ) \over {{\omega'' -\omega }} }
t (\vec k , \vec k'' ; \omega'') t^{*} (\vec k' , \vec k'' ; \omega'') 
\cr &-&  {\cal P} \int d^3 k''{  (\sqrt{{\omega''} ^2 +p^2} -\omega'' ) \over {\omega'' -\omega' } } 
t (\vec k , \vec k'' ; \omega'') t^{*} (\vec k' , \vec k'' ; \omega'') \}
\label{a1}
\end{eqnarray} 
is also real. This term contains  
the complex expression  $t(\vec k, \vec k'' ;\omega'')
t^*(\vec k', \vec k'' ;\omega'')$, which we will have to rewrite in order to show that the
integration over it results in a real number. First, we note that only the half-shell t-matrix
enters the integration in Eq.~(\ref{a1}). Via the Heitler equation it can be related to the
K-matrix,
\begin{eqnarray} 
t(\vec k, \vec k'' ;\omega'')&=&K(\vec k, \vec k'' ;\omega'')
- i \pi \sqrt{ {k''} ^2 + m ^2} \: k'' \int d \hat{k}''' 
K (\vec k, \vec k''' ; \omega'')  t(\vec k''', \vec k'' ;\omega''),
\label{a3}
\end{eqnarray}
where $\vert \vec k'''  \vert = k'' $ is on-energy-shell.
The K-matrix is real and defined in the standard fashion as 
\begin{eqnarray} 
K (\vec k, \vec k'' ; \omega'') = v(\vec k, \vec k'')
+ {\cal P} \int  d^3 k'''  { v(\vec k, \vec k''')  K (\vec k''', \vec k'' ; \omega'') 
 \over {{\omega'' -\omega''' }} }  
\label{a2}
\end{eqnarray}
In order to carry out the angular integration in Eq.~(\ref{a2}), 
we use the partial wave representations of the t- and K-matrix,
\begin{eqnarray} 
t(\vec k, \vec k'' ;\omega'')&=& \sum _{l m  }  t_l  (k,  k''
;\omega'') Y_{l m  } (\hat k) Y^*  _{l m  } (\hat k''),
\label{a4}
\end{eqnarray}
and 
\begin{eqnarray} 
K(\vec k, \vec k'' ;\omega'')&=& \sum _{l m  }  K_l  (k,  k''
;\omega'') Y_{l m  } (\hat k) Y^*  _{l m  } (\hat k'').
\label{a5}
\end{eqnarray}
Inserting these partial wave expressions into Eq.~(\ref{a2}) leads to a partial wave
representation of $t$ as
\begin{equation}
 t_l (k, k'';\omega'') = K_l (k, k'';\omega'') 
  ( 1- i \pi \sqrt{ {k''} ^2 + m ^2} \; k'' \;  t_l(k'', k'';\omega'')).
\end{equation}
Thus, 
the half-shell t-matrix $t_l (\vec k, \vec {k''} ; \omega '' )$ receives its complex parts 
only from the factor $
 ( 1- i \pi \sqrt{ {k''} ^2 + m ^2} k''  t_l ( k'',  k'';\omega''))$ which does not depend on $k$. 
Using the partial wave expansions of Eqs.~(\ref{a4}) and (\ref{a5}) leads to
\begin{eqnarray} 
\label{eq:a6}
\lefteqn{ \int d \hat{k}'' \;  t (\vec k, \vec k'' ; \omega'')  t^* (\vec k',\vec k'';\omega'')} \\
 &=&\sum _{l m}  t_l (k,k'';\omega'')  t_l^*  (k', k'';\omega'') 
 Y_{l m  } (\hat k) Y^*  _{l m  } (\hat k') \nonumber \\
%\cr &=&\sum _{l m  }  K_l  (k,  k''
%;\omega'')  K_l  (k',  k''
%;\omega'')  \cr 
%& \times&
%( 1- i \pi \sqrt{ {k''} ^2 + m ^2} k''  t_l ( k'',  k''
% ;\omega'') ) ( 1+ i \pi \sqrt{ {k''} ^2 + m ^2} k''  t_l^* ( k'',  k''
% ;\omega'') )
% Y_{l m  } (\hat k) Y^*  _{l m  } (\hat k')
 &=&\sum _{l} (2l +1)  K_l  (k, k'';\omega'')  K_l (k', k'';\omega'') 
\left( 1 +\pi ^2  ({ {k''} ^2 + m ^2})  {k'' }^2  \vert   t_l ( k'',  k''
 ;\omega'') \vert ^2  \right) P_l( \hat  k \cdot \hat {k'} ).  \nonumber
\end{eqnarray}
Here $P_l$ is the Legendre polynomial.  
The expression given in Eq.~(\ref{eq:a6}) is manifestly real 
and consequently the expression given in Eq.~(\ref{a1}) is real.

\section{Partial Wave Representation }

In a partial wave representation 
the relativistic Faddeev Eq. (\ref{Faddeevrel}) is explicitly given  as
\begin{eqnarray}
\phi_\alpha ( k, p)&=&  { 1 \over E_b - {\cal E } (  k ,   p) }  
\sum _ {\alpha ' \alpha '' } 
\int  _ {0 }^\infty d p ' { p '}^2 
\int _ {-1 }^1 d x { T_{\alpha \alpha'} (k,\kappa_1; p') \over {{\kappa_1}^{l'}}}
\cr & & \times {  G _ { \alpha ' \alpha ''  } ( p , p ', x) \over { 
     {\cal N }_ 1  ( p, p', x ){\cal N }_ 2
( p, p', x )  }   }  
 { \phi_{\alpha '' } 
 ( \kappa_2 ,   {p'} ) \over { \kappa_2} ^ {l''}  }   
\label{Faddeevrelpartial}
\end{eqnarray}
where 
\begin{eqnarray}
 G _ { \alpha  \alpha '  } ( p , p ', x) = \sum _ {\cal L  } 
P  _ {\cal L  }  ( x)  \sum _ { l_ 1 + l_ 2 = l     }  
 \sum _ { l'_ 1 + l'_ 2 = l'    }  
{ \left\{ ( 1 + y_ 1 )   p \right\}  } ^{  l_ 2 +  l'_ 2   }  
{  \left\{ ( 1 + y_ 2)  p '  \right\}  } ^{  l_ 1 +  l'_ 1  } 
g  _ {\alpha \alpha '   }  ^ { {\cal L  }  l_ 1   l_ 2   l'_ 1  
l'_ 2   },
\end{eqnarray}
and
\begin{eqnarray}
 \kappa_1 &=& \sqrt{ {p'} ^2 + {(1+y_ 1)^2  \over 4} { p} ^2 + (1+y_ 1)  p p'
  x  }, \cr 
 \kappa_2 &=& \sqrt{ {p} ^2 + {(1+y_ 2)^2  \over 4} {p'} ^2 + (1+y_ 2)  p p'
  x  },
\end{eqnarray}
and
\begin{eqnarray}
y_1 &=& y_1 (p,p',x) \\ 
 &=&
 { \sqrt{ m^2 +{p' }^2   } -  \sqrt{m^2+ p^2 +{p' }^2 +2 p p' x } \over {
\sqrt{ m^2 +{p' }^2   } + \sqrt{m^2+ p^2 +{p' }^2 +2 p p' x  } + \sqrt{
( \sqrt{ m^2 +{p' }^2   } +  \sqrt{m^2+ p^2 +{p' }^2 +2 p p' x  } )^2 - p^2 }}},
\cr  & & \cr
y_2&=&  y_1 (p',p,x). \nonumber
\end{eqnarray}
When explicitly calculating the normalization factor ${\cal N}({{\vec p}'},-{\vec p}-{{\vec p}'})$, 
it turns out that ${\cal N}({{\vec p}'},-{\vec p}-{{\vec p}'}) \rightarrow {\cal N }_1
(p,p',x)$ with 
\begin{eqnarray}
\lefteqn{ {\cal N }_ 1  ( p, p', x ) =} \\
& & \left( 4 \sqrt{m^2+{p'}  ^2} \sqrt{m^2+
p^2 +{p' }^2 +2 p p' x } \over { \sqrt{ ( \sqrt{m^2+{p'} ^2 } +  \sqrt{m^2+
p^2 +{p' }^2 +2 p p' x })^2 - p ^2  }  ( \sqrt{m^2+{p'} ^2 } +  \sqrt{m^2+
p^2 +{p' }^2 +2 p p' x })  } \right)^{1/2}. \nonumber 
\end{eqnarray}
In a similar vain, ${\cal N}(-{\vec p}-{\vec p'}, {\vec p}) \rightarrow {\cal N }_2 ( p,
p', x )$, with 
\begin{equation}
 {\cal N }_ 2  ( p, p', x ) =  {\cal N }_ 1 ( p', p, x ) .
\end{equation}
The index $\alpha$ summarizes a  set of quantum numbers (channels)
\begin{eqnarray}
\vert \alpha \rangle = 
\vert (ls)j (\lambda {1\over 2 }) I (j I) J (t {1 \over
2})T \rangle , 
\end{eqnarray}
where 
$l,s,j$ and 
$ t $ are orbital angular momentum, total  spin, total angular momentum j and total 
isospin in the two-body subsystem. The indices  
$\lambda, I $, $ J  $, and $ T$  stand for  the orbital angular momentum,  the 
total angular momentum of the third particle, the total three-body angular momentum,
and the total isospin \cite{Report,TEXT}). The quantity
 $g_{\alpha \alpha '}^ {{\cal L} l_1 l_2 l'_1   l'_2}$  represents
the standard  permutation operator coefficient.
  
Finally, when taking the  limits
\begin{eqnarray}
y_1 , y_2 &\to& 0 , \cr
 {\cal N }_1 , {\cal N  }_2 &\to& 1,
\end{eqnarray}
one obtains the nonrelativistic result
\begin{eqnarray} 
 {\cal E } (k,p) &\to& { k^2 \over m  }+ { 3 p^2 \over 4 m  } ,\cr
T &\to& t^{(nr) },
\end{eqnarray}
and the relativistic Faddeev equation, Eq.~(\ref{Faddeevrel}),
 reduces to the nonrelativistic one.

%\pagebreak
%%%%%%%%%%%%%%%%%%%%%%%%%%%%%%%%%%%%%%%%%%%%%%%%%%%%%%

\begin{table}[tp]
\begin{center}
\begin{tabular}[t]{l|r|r|r|r}
interaction           & $E_{b}$ &  $E_{b} ^{(nr)}$ & $\Delta$  &  
$E_{b} \; ({\cal N} \to 1$) \cr
\hline
RSC \cite{RSC}   &  -6.59  &   -7.02\cite{EB}  &   0.43 & -6.63 \cr
%RSC  $^*$  \cite{Gl86}    &  -6.67  &        -7.02      &   0.35 & -6.71  \cr
CD-Bonn \cite{CDBONN} &  -7.98  &   -8.33     &  0.35   & -8.03 \cr
Nijmegen II\cite{nijm93}   & -7.22  & -7.65   &  0.43     & -7.27 \cr
  Nijmegen I\cite{nijm93}   & -7.71 & -8.00   &   0.29    & -7.76 \cr
Nijmegen 93 \cite{nijm93}   & -7.46 & -7.76   & 0.30      & -7.51 \cr
AV18 \cite{av18}   &  -7.23       &  -7.66  & 0.43      & -7.27 \cr
 Yamaguchi I \cite{Gibson}       &   -9.93     &  -10.13 &  0.20   &  -10.04      \cr
 Yamaguchi II\cite{Gibson}        & -8.30       &  -8.48 &  0.18   &  -8.40 \cr
exp.                &   -8.48     &            &           &     \cr
    \end{tabular}
    \caption{ The relativistic ($E_b$) and nonrelativistic ($E_b^{(nr)}$) 
triton binding energies in MeV obtained from different nonrelativistic
potentials.  The quantity $\Delta$ is defined as $\Delta \equiv E_b - E^{(nr)}_b $.
For comparision we also list the results of the relativistic calculation when the 
Jacobian function ${\cal N}$ is set to 1. } 
\label{table1}
\end{center}
\end{table}

\pagebreak
%%%%%%%%%%%%%%%%%%%%%%%%%%%%%%%%%%%%%%%%%%%%%%%%%%%%%%

%\begin{figure}[hbt]
%\psfig{file=flowchart.eps,angle=0}%,scale=0.7}
%\caption{Flow-chart of our calculation together with the equation numbers. 
%The dashed flow was followed in \protect \cite{Gl86}, see especially Eq. (3.27) therein.
%}
%\label{flowchart}
%\end{figure}

\begin{thebibliography}{100}

\bibitem{CDBONN}  
{R. Machleidt} {\it et al.},
Phys. Rev. C {\bf {53}}, R1483 ({1996}).
\bibitem {nijm93}
{{V. G. J. Stoks} {\it et al.}},
 {Phys. Rev. C} {\bf {49}}, {2950} ({1994}).

\bibitem{av18}{{R. B. Wiringa}} {\it et al.}, Phys. Rev. C {\bf 51}, 38
(1995).

\bibitem{Friar} J. L. Friar, {\it et al.}, Phys. Lett. { \bf  B 311}, 4 (1993).


\bibitem{anbench} A.~Nogga, D.~H\"uber, H.~Kamada, W.~Gl\"ockle,
Phys. Lett. {\bf B 409}, 19 (1997).
\bibitem{NoggaPhd} A. Nogga, PhD Thesis, Ruhr-University Bochum, 2001.

\bibitem{Coester1} F. Coester, Proceedings to the 11th International Conference on
Recent Progress in Many-Body Theories, nucl-th/0111025.
\bibitem{Rupp}
G. Rupp, J. A. Tjon, Phys. Rev. C {\bf 45}, 2133 (1992). 

\bibitem{Sammarruca}
F. Sammarruca, R. Machleidt, Few-Body Syst.  {\bf 24}, 87 (1998).

\bibitem{Stadler}
A. Stadler, F. Gross, M.  Frank,  Phys. Rev. C {\bf 56}, 2396 (1997);
A. Stadler, F. Gross, Phys. Rev. Lett. {\bf 78}, 26 (1997).

\bibitem{Bakamjian}
B. Bakamjian, L. H. Thomas, Phys. Rev. {\bf 92}, 1300 (1952).
\bibitem{Foldy}
L. L. Foldy, Phys. Rev. {\bf 122}, 275 (1961); R. A. Krajcik,
L. L. Foldy, Phys. Rev. D {\bf 10}, 1777 (1974).
\bibitem{Keister} 
B. D. Keister, AIP Conf. Proc. {\bf 334}, 164 (1995);
B. D. Keister, W. N. Polyzou, Adv. Nucl. Phys. {\bf 20 }, 225 (1991).

\bibitem{Carlson}
J. Carlson, V. R. Pandharipande, R. Schiavilla, Phys. Rev. C{\bf 47},
484 (1993); J. L. Forest, V. R. Pandharipande, J. L. Friar,
Phys. Rev. C {\bf 52 }, 568 (1995); J. L. Forest, V. R. Pandharipande, 
J. Carlson, R. Schiavilla, Phys. Rev. C {\bf 52}, 576 (1995).

\bibitem{Gl86}
W. Gl\"ockle, T-S. H. Lee, and F. Coester, Phys. Rev. C {\bf 33}, 709
(1986); F. Coester, Helv. Phys. Acta {\bf 38}, 7 (1965);
 L. M\"uller, Nucl. Phys. {\bf A360}, 331 (1981); 
W. Gl\"ockle, L. M\"uller, Phys. Rev. C {\bf 23}, 1183 (1981).

\bibitem{Kamada}
H. Kamada and W. Gl\"ockle, Phys. Rev. Lett. {\bf 80}, 2547 (1998).

\bibitem{Polyzou} T.W. Allen, G.L. Payne, and Wayne N. Polyzou, 
Phys. Rev. C {\bf 62}, 054002 (2000).

\bibitem{RSC} R. V. Reid, Ann. Phys. {\bf 50} , 411 (1968).

\bibitem{Report}
 W. Gl\"ockle, H. Wita\l a, D. H\"uber, H. Kamada, J. Golak,
Phys. Rep. {\bf 274}, 107 (1996).

\bibitem{TEXT}
 W. Gl\"ockle, {\it The Quantum-Mechanical Few-Body Problem} (Springer
 Verlag, Berlin, Heiderberg, 1983).

\bibitem{Gibson}
B. F. Gibson, D. R. Lehman,  Phys. Rev. C {\bf 14}, 685 (1976).

\bibitem{Abfalterer}
W. P. Abfalterer {\it et al.}, Phys. Rev. Lett. {\bf 81}, 57 (1998).
\bibitem{Witala99}
H. Wita\l a  {\it et al.}, Phys. Rev. C{\bf 59}, 3035 (1999).

\bibitem{Kamada2000}
H. Kamada, Few-Body Syst. Suppl.   {\bf 12 }, 433 (2000). 

\bibitem{EB} W. Gl\"ockle, Nucl. Phys. {\bf A 381}, 343 (1982);
C. Hadjuk, P. U. Sauer, Nucl. Phys. { \bf A 369}, 321 (1981);
I. R. Afnan, N. D. Birrell, Phys. Rev. C {\bf 16}, 823 (1977).


\end{thebibliography}
\end{document}